# Wigner time for electromagnetic radiation in plasma

S. V. Gaponenko[1] and D. V. Novitsky[2]

B. I. Stepanov Institute of Physics, National Academy of Sciences of Belarus,
68 Nezavisimosti Ave., Minsk 220072, Belarus

**Abstract**: Wave tunneling is an intriguing phenomenon spanning different branches of physics, from quantum mechanics to classical electrodynamics and optics. The Wigner (or phase) time is proved to be an adequate measure to describe wave transit through a potential barrier or material layer in the tunneling regime. Here, we analytically and numerically calculate the Wigner time for electromagnetic-radiation propagation through the layer of both lossless and lossy plasmas. It is shown that the plasma frequency is the key parameter governing the value of Wigner time allowing us to interpret tunneling as due to the reaction of plasma as a whole. We analyze the Wigner time for obliquely incident waves of TE and TM polarizations and discuss the meaning of negative Wigner times appearing in the lossy case in the low-frequency range and close to the plasma frequency. The results show that plasma deserves attention as a perspective object for tunneling studies.

## I. Introduction

In optics and radiophysics propagation of electromagnetic radiation through a finite layer of a medium by means of evanescent field represents a classical counterpart to quantum-mechanical tunneling with the length of the medium layer supporting wave propagation solely via evanescence being an analog of a potential barrier width in quantum mechanics. In this context, the electromagnetic analogs of the Hartman paradox have become the subject of extensive analysis. The essence of this paradox is the asymptotic independence of the Wigner time (phase time) for an electron and electromagnetic radiation on the barrier length (width) $L$ in the limit of extremely low transmittivity of the barrier under consideration [1, 2]. The Wigner time is defined as

$$\tau_\varphi = \frac{d\varphi}{d\omega}, \qquad (1)$$

where $\omega$ is the radiation frequency and $\varphi$ is the phase of the complex transmission coefficient, $t$. This notion has been introduced by E. Wigner in 1955 [3] as a measure for scattering events in quantum physics; it was further proved to be a one-dimensional approximation of the lifetime matrix [4]. Its independence on the barrier length (width) indicates at first glance a possibility of seemingly superluminal propagation since tunneling time defined by Eq. (1) may become shorter than the physical time $t = L/c$, where $c$ is the speed of light in vacuum. Thorough analysis performed in several studies (see [5-8] and references therein) have led to the following conclusions: (i) speed may not be the relevant term in tunneling, since there is no well-defined physical object whose motion is traceable within the barrier region; (ii) very low energy transmission coefficient defines very low rate of energy transfer whenever $\tau_\varphi < L/c$ holds; (iii) the Wigner time, when applied to tunneling, does not define signaling rate. Certain criticism has been expressed on the very attempt to analyze time-dependent tunneling events using time-independent equations [9, 10]. However, comparison of Wigner times with the correct results based on wave packets tunneling instead of plane waves [11] has shown very good agreement indicating that tunneling analysis in terms of phase time is meaningful. Energy storage and subsequent release in the barrier region with the relevant lifetime instead of transit phenomena have also been suggested to characterize tunneling rate [12]. In addition to wave equation analysis, photon tunneling has become a subject of discussions up to bringing about virtual photons [13] and tachyons [14].

---

[1] s.gaponenko@ifanbel.bas-net.by
[2] Corresponding author: dvnovitsky@gmail.com

The typical model structures for electromagnetic tunneling are a couple of prisms with a spacer where tunneling occurs by means of frustrated total internal reflection; a photonic band gap material, e. g., a multilayer stack in which evanescent modes develop for radiation frequencies within the spectral gap (photonic stop-band) range; and an undersized waveguide. There is also a certain tendency that the Wigner time for the above structures is defined by the inverse radiation frequency $1/\omega$ as has been for the first time highlighted by Haibel and Nimtz [15] and then supported to large extent by numerical simulations [16]. The complex generalization of the Wigner time was introduced recently [17] to characterize light interaction with lossy chaotic systems having poles and zeros in their scattering matrices.

Yet another case of electromagnetic radiation propagation via evanescent waves is transparency of thin metal films in optics (optical response of metals can be reasonably described in terms of dense electron gas) and plasma layers in a more general context including radiofrequencies and microwaves. Evanescence becomes the only possible way for radiation penetration through a plasma layer in the range of negative permittivity, ε<0. In this context, metal-film-covered glass slides used as sunglasses represent a genuine electromagnetic analogy to quantum-mechanical tunneling which becomes apparent when the time-independent one-dimensional Schrödinger equation is compared to the Helmholtz equation in electrodynamics. Notably, this straightforward analogy has been outlined already by W. Heisenberg at the very dawn of quantum mechanics [18].

In this paper, we analyze the Wigner time for light propagating through plasma which was overlooked in previous publications on tunneling. We start with the brief discussion of the ideal-plasma case studied in detail recently [19]. However, real media always feature certain losses and those preliminary analysis of the Wigner time for electromagnetic tunneling through plasmas is to be complemented by the analysis taking loss into account. Here, we fill this gap calculating the Wigner time for light tunneling through non-ideal plasma. In contrast to the case of ideal plasma considered in Ref. [19], introduction of loss results in the appearance of additional features such as negative Wigner times in the low-frequency range and close to the plasma frequency. We also discuss the peculiarities of Wigner time for different light polarizations appearing under oblique incidence. Our approach allows us to analyze the tunneling phenomena in realistic plasma media.

**II. Tunneling through ideal plasma**

In an ideal plasma, the permittivity obeys (only electron contribution is considered, since the masses of ions are much larger than the electron one) the known Drude relation

$$\varepsilon(\omega) = 1 - \frac{\omega_p^2}{\omega^2}, \quad \omega_p^2 = \frac{Ne^2}{m\varepsilon_0} \quad , \qquad (2)$$

where $N$ is the electron concentration, $e$ is the elementary charge, $m$ is the electron mass, and $\varepsilon_0$ is the vacuum permittivity. It was shown that in the case of an ideal lossless plasma, the Wigner time in the low-frequency limit $\omega \ll \omega_p$ asymptotically tends to $2/\omega_p$ for a plasma layer in vacuum and to $2n/\omega_p$ for a plasma layer in an ambient lossless dielectric with refractive index $n$ [7, 19]. The relevant equation obtained under the low-transmission assumption reads

$$\tau_\varphi = \frac{2}{\sqrt{\omega_p^2 - \omega^2}} \frac{n}{1 + (n^2-1)\frac{\omega^2}{\omega_p^2}} \tag{3}$$

and reduces for $n = 1$ to

$$\tau_\varphi = \frac{2}{\sqrt{\omega_p^2 - \omega^2}}. \tag{4}$$

The low-frequency ($\omega \ll \omega_p$) asymptotic $2/\omega_p$ value found for the Wigner time in the case of a lossless plasma layer suggests that the very tunneling event in this case may be possibly interpreted as a splash or flinching of a plasma layer as a whole, since the plasma frequency defines the extreme rate of plasma response to electromagnetic stimuli. For a quick reference, it is reasonable to recall plasma frequency values for a number of representative cases. In typical metals, $\omega_p$ is of the order of $10^{16}$ Hz; in semiconductors at high optical excitation, the non-equilibrium electron-hole plasma features the plasma frequency in the range of $10^{10}…10^{13}$ Hz; for a laboratory plasma (e.g., gas discharge), it is of the order of $10^{10}$ Hz; for the ionosphere, the plasma frequency of the order of $10^8$ Hz gives rise to short radio waves reflection enabling long-distance communication.

**III. Tunneling through lossy plasma**

In this Section, we consider the simplest generalization of Eq. (2) taking the permittivity to be

$$\varepsilon(\omega) = 1 - \frac{\omega_p^2}{\omega^2 + i\gamma\omega}, \tag{5}$$

where $\gamma$ is the damping rate introducing loss into our model. For simplicity, the ambient is assumed to be air with the permittivity 1; generalization to the arbitrary environment is straightforward. Radiation is incident on the plasma layer of thickness $L$ at an angle $\theta$, so that the calculations are generally dependent on light polarization. The expression for the complex transmission coefficient can be found in standard textbooks on optics, e.g., in Ref. [20]. It is convenient to write it as follows

$$t = \frac{1}{\cos\beta - i\alpha\sin\beta} = |t|e^{i\varphi}, \tag{6}$$

where $\alpha = \frac{p^2(\omega) + \cos^2\theta}{2p(\omega)\cos\theta} = \alpha' + i\alpha''$ and $\beta = \frac{\omega}{c}L\sqrt{\varepsilon(\omega) - \sin^2\theta} = \beta' + i\beta''$. Auxiliary function $p(\omega)$ has different forms depending on polarization: for TE waves, it is $p_{TE}(\omega) = \sqrt{\varepsilon(\omega) - \sin^2\theta}$, whereas for TM waves, we have $p_{TM}(\omega) = \sqrt{\varepsilon(\omega) - \sin^2\theta}/\varepsilon(\omega)$.

Appearance of the tunneling regime can be assessed by considering the parameter $\beta$ containing information on the phase and damping of radiation. For a lossless material and normal incidence, evanescent waves occur at $\varepsilon(\omega) < 0$, when $\beta = \frac{\omega}{c}L\sqrt{\varepsilon(\omega)} = i\beta''$ is purely imaginary. This is the classic case of tunneling. For the plasma with permittivity given by Eq. (2), this occurs at frequencies $\omega < \omega_p$. In the presence of losses, the analogy between tunneling and evanescence cannot be strictly proved, since the waves are now propagating. However, for low losses ($\gamma \ll \omega_p$) and non-normal incidence, the direct correspondence with the classic tunneling can be established, when the root expression in $\beta$ is negative, i.e., $\varepsilon(\omega) - \sin^2\theta \approx \cos^2\theta - \frac{\omega_p^2}{\omega^2} < 0$. This condition is satisfied for the low enough frequencies in the range depending on the incidence angle, $\omega < \omega_p/\cos\theta$. We further consider this case of tunneling-like response.

The Wigner time (1) can be easily computed numerically for the phase extracted from Eq. (6). In general, it depends on the layer thickness via the parameter $\beta$. However, we can obtain an analytical estimate for $\tau_\varphi$ neglecting dependence on $L$. Such an estimate corresponds to the conditions for the Hartman paradox and, therefore, deserves our attention. To simplify the expression for $\beta$, we use the same assumptions as above, $\gamma \ll \omega_p$ and $\omega < \omega_p/\cos\theta$. Then, for the imaginary part, we have $\beta'' \approx \frac{\omega_p}{c} L \sqrt{1 - \frac{\omega^2}{\omega_p^2}\cos^2\theta} \gg 1$ that corresponds to the low-transmission limit widely used in the tunneling literature [16]. This condition is satisfied for thick enough layers and not too close to the limiting frequency $\omega_p/\cos\theta$. The real part can be written as $\beta' \approx \left(\frac{\omega_p}{c}L\right)^2 \frac{\gamma}{\omega} \frac{1}{\beta''} \ll 1$, so that the influence of $L$ on the phase is neglected. Then, we obtain the approximate expression for the phase as follows

$$\varphi_t \approx -\arctan\frac{\alpha''}{1+\alpha'}. \qquad (7)$$

To calculate the Wigner time, we insert Eq. (7) into Eq. (1) and choose the proper (for given polarization) function $p(\omega)$ needed for evaluation of $\alpha$ as described after Eq. (6). The resulting expressions are too cumbersome to be written analytically, so it is convenient to present the Wigner time as the series expansion in powers of the small damping parameter $\gamma/\omega_p$ (this can be readily done using a computer algebra system). Leaving only the lowest-order terms, we have

$$\tau_\varphi^{TE}\omega_p = \frac{2\cos\theta}{\sqrt{1-\frac{\omega^2}{\omega_p^2}\cos^2\theta}} + \frac{\gamma}{\omega_p}\frac{-1+\frac{\omega^2}{\omega_p^2}\cos^2\theta - 2\frac{\omega^4}{\omega_p^4}\cos^4\theta}{2\frac{\omega^2}{\omega_p^2}\left(1-\frac{\omega^2}{\omega_p^2}\cos^2\theta\right)^2} + O(\gamma^2/\omega_p^2), \qquad (8)$$

$$\tau_\varphi^{TM}\omega_p = \frac{4\cos\theta\left(1-\frac{\omega^2}{\omega_p^2}\cos 2\theta\right)}{\sqrt{1-\frac{\omega^2}{\omega_p^2}\cos^2\theta}\left[1+\left(1-2\frac{\omega^2}{\omega_p^2}\right)\cos 2\theta\right]} + O(\gamma/\omega_p). \qquad (9)$$

In Eq. (9), we have left only the first term of the series to be more concise. It is easy to see that these formulas are reduced to Eq. (4) for the normal incidence and ideal plasma ($\theta = 0, \gamma = 0$). The case of normal incidence (in the frequency range $0 < \omega < \omega_p$) is convenient to analyze the influence of losses. The dependence according to Eq. (4) is shown in Fig. 1 with the black line. It is seen that introducing small losses leaves the dependence mostly unchanged with the exception of narrow range close to zero frequency where the Wigner time can take even negative values. The range of negative $\tau_\varphi$ gets wider with increasing $\gamma$.

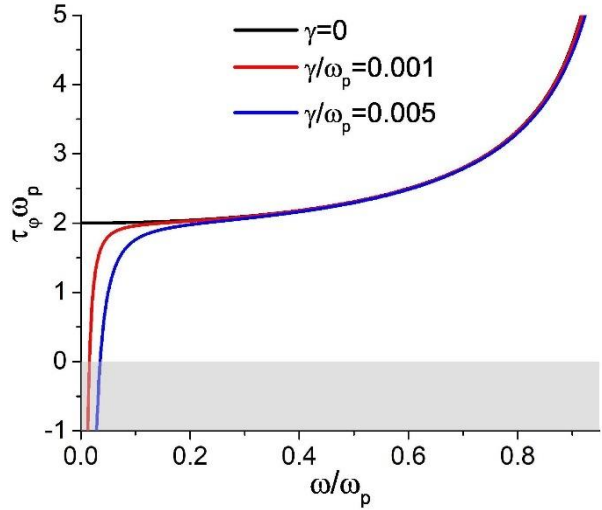

Fig. 1. Frequency dependence of the Wigner time for normally incident radiation.

Let us discuss the applicability of formulas (8) and (9) comparing them with the numerical calculations with the unsimplified expression (6). The results of such a comparison is shown in Fig. 2 for $\theta = \pi/3$, $\gamma/\omega_p = 0.01$ and different layer thicknesses. One can see that the general features of Wigner time behavior are adequately described by the approximate formulas in the full frequency range $0 < \omega < \omega_p/\cos\theta = 2\omega_p$. The difference between the approximate and exact results grows with increasing $L$ and also with approaching the boundaries of the frequency range. For example, approaching $\omega_p/\cos\theta$, the approximate expressions give negative values, whereas the exact values grow indefinitely. Far from these boundaries, the Wigner time almost does not depend on the thickness that is especially evident for TE waves.

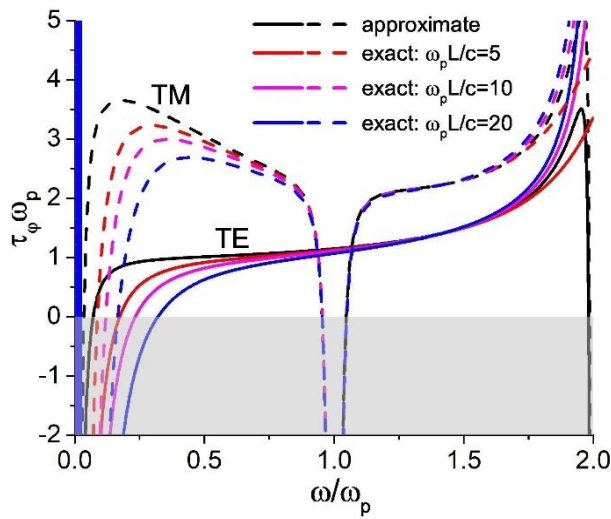

Fig. 2. Frequency dependence of the Wigner time for $\theta = \pi/3$ and $\gamma/\omega_p = 0.01$ calculated with the approximate and exact expressions. Solid lines are for the TE waves, dashed lines are for the TM waves.

Figure 3 shows how the Wigner time computed with Eqs. (8) and (9) depends on the main parameters of the problem. First, we fix the incidence angle ($\theta = \pi/3$) and analyze the influence of the damping rate [Fig. 3(a)]. One can see that $\tau_\varphi$ weakly depends on $\gamma$ for TE waves, whereas for the TM waves, $\tau_\varphi$ decreases with $\gamma$ and mostly $\tau_\varphi^{TM} > \tau_\varphi^{TE}$. Second, we fix the damping rate ($\gamma/\omega_p = 0.05$) and analyze the Wigner time for different incidence angles [Fig. 3(b)]. We see the widening of the frequency range $[0, \omega_p/\cos\theta]$ with growing angle. For TE waves, $\tau_\varphi$ clearly shortens from $\tau_\varphi^{TE} \approx 2/\omega_p$ at $\theta = 0$ to $\tau_\varphi^{TE} \approx 1/\omega_p$ at $\theta = \pi/3$. For TM waves, on the contrary, $\tau_\varphi^{TM}$ for nonzero $\theta$ is mostly larger than in the case of normal incidence. The large values of $\tau_\varphi^{TM}$ above the plasma frequency at low incidence angles is due to the rapid growth of curves as we approach the frequency $\omega_p/\cos\theta$.

A characteristic feature of the obtained solutions for radiation tunneling through the lossy plasma is the appearance of negative values of the phase time (the corresponding regions in figures above are shaded in gray). Negative values of $\tau_\varphi$ appear in all cases (TE and TM waves, normal or oblique incidence) with increasing losses at low frequencies $\omega \ll \omega_p$. Appearance of $\tau_\varphi < 0$ can be easily demonstrated in Eq. (8) where the second term becomes an infinitely growing negative number in the low-frequency limit, $\omega \rightarrow 0$. The negative time at $\omega_p/\cos\theta$ appears for both polarizations only in the approximate formulas and is absent in the exact calculations as seen in Fig. 2. For the TM waves under oblique incidence and for nonzero losses, the negative time occurs also in the vicinity of plasma frequency that is clearly associated with plasmon excitation by TM-polarized radiation. Negative values of the Wigner time have been encountered before [21] but have not received unambiguous physical interpretation. In fact, the possibility of negative $\tau_\varphi$ was mentioned already by Wigner [3] and was connected to the regions of phase decreasing with energy far from the scattering resonances. The negative Wigner time resulting in unphysical negative density of modes was reported for the pulses traversing photonic crystals and was interpreted as due to proximity to the absorption line [11]. In our case, the negative values of $\tau_\varphi$ also occur near the specific resonant or singular points ($\omega = 0$ and $\omega = \omega_p$). We believe that the negative values of the tunneling time are devoid of physical meaning and indicate a possible going beyond the applicability limits of the approach used and the impossibility of using the concept of phase time under such conditions. Therefore, it is necessary to use and interpret the results with caution when the problem parameters approach the intervals for which the Wigner time becomes negative.

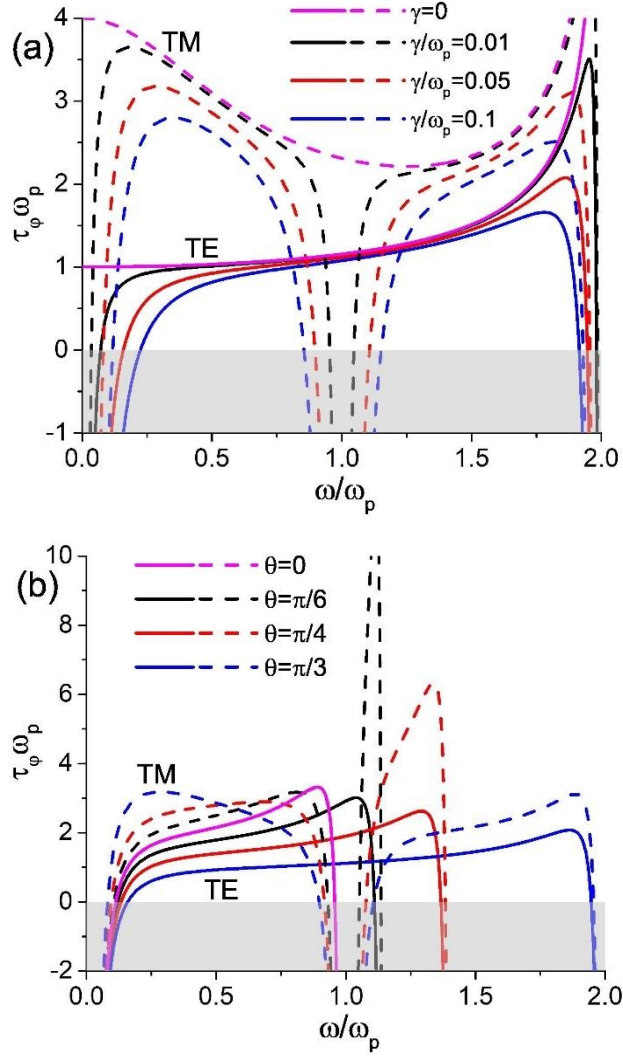

Fig. 3. Frequency dependence of the Wigner time for (a) different losses at $\theta = \pi/3$ and (b) different angles at $\gamma/\omega_p = 0.05$. Solid lines are for the TE waves, dashed lines are for the TM waves.

**IV. Conclusion**

In conclusion, we have studied behavior of the Wigner time $\tau_\varphi$ for electromagnetic radiation tunneling through a non-ideal (lossy) plasma layer. The approximate relations for $\tau_\varphi$ are obtained, which, in the limit of normal incidence and zero losses, give a known low-frequency asymptotic $\tau_\varphi \to 2\omega_p^{-1}$. Unlike tunneling through the dielectric structures (prisms with frustrated total internal reflection, photonic crystals, undersized waveguides), for which the quantity that specifies the Wigner tunneling time in the limit of low transmittance (wide and high barrier) is the inverse frequency of the propagating radiation, the Wigner time for the lossless and low-loss plasmas is determined by the inverse plasma frequency, $\tau_\varphi \sim \omega_p^{-1}$, and does not depend on the radiation frequency. For obliquely incident radiation, the results differ depending on light polarization. For TE waves, the result $\tau_\varphi \sim \omega_p^{-1}$ holds almost in the entire low-frequency range ($\omega < \omega_p/\cos\theta$) with $\tau_\varphi$ decreasing with the incidence angle. For TM waves, an additional singularity appears in the dependence $\tau_\varphi(\omega)$ near the plasma frequency. The obtained results can be used to estimate the light-pulse transit times through realistic plasmas with the exception of singular points vicinity where the phase time can take on unphysical negative values. We believe that our approach clarifies the physics of light tunneling as applied to a rather simple plasma model.